*Title:* **Investigations of Neutron Characteristics for Salt Blanket Models; Integral Fission Cross Section Measurements of Neptunium, Plutonium, Americium and Curium Isotopes**

*Author(s):* Eduard FOMUSHKIN, Gleb NOVOSELOV, Vyacheslav GAVRILOV, Mikhail KUVSHINOV, Vladimir BOGDANOV, Georgij MASLOV, Vladimir VYACHIN, Vladimir GORELOV, Victor EGOROV, Vladimir IL'IN, Dmitrij PESHEKHONOV, Alexander SHVETSOV, Yurij TITARENKO, Vyacheslav KONEV, Mikhail IGUMNOV, Vyacheslav BATYAEV, Evgenij KARPIKHIN, Valerij ZHIVUN, Alexander KOLDOBSKY, Ruslan MULAMBETOV, Dmitrij FISCHENKO, Svetlana KVASOVA, Alexander LOPATKIN, Victor MURATOV, Anatolij LOSITSKY, Boris KURUSHIN, Hideshi YASUDA and Stepan MASHNIK

# Investigations of Neutron Characteristics for Salt Blanket Models; Integral Fission Cross Section Measurements of Neptunium, Plutonium, Americium and Curium Isotopes

Eduard FOMUSHKIN[1,*], Gleb NOVOSELOV[1], Vyacheslav GAVRILOV[1], Mikhail KUVSHINOV[1], Vladimir BOGDANOV[1], Georgij MASLOV[1], Vladimir VYACHIN[1], Vladimir GORELOV[1], Victor EGOROV[1], Vladimir IL'IN[1], Dmitrij PESHEKHONOV[1], Alexander SHVETSOV[1], Yurij TITARENKO[2], Vyacheslav KONEV[2], Mikhail IGUMNOV[2], Vyacheslav BATYAEV[2], Evgenij KARPIKHIN[2], Valerij ZHIVUN[2], Alexander KOLDOBSKY[2], Ruslan MULAMBETOV[2], Dmitrij FISCHENKO[2], Svetlana KVASOVA[2], Alexander LOPATKIN[3], Victor MURATOV[3], Anatolij LOSITSKY[4], Boris KURUSHIN[4], Hideshi YASUDA[5] and Stepan MASHNIK[6]

[1]*Russian Federal Nuclear Center – VNIIEF, 607188, Sarov (Arzamas-16), Nizhny Novgorod region, Russia*
[2]*Institute for Theoretical and Experimental Physics, 117259, Moscow, Russia*
[3]*Research and Development Institute of Power Engineering, POB 785, 101000, Moscow, Russia*
[4]*JSC Chepetsky Mechanical Plant, 427600, Glazov, Udmurt Republic, Russia*
[5]*Japan Atomic Energy Research Institute, Tokai, Ibaraki, 319-1195, Japan*
[6]*Los Alamos National Laboratory, Los Alamos, NM, 87545, USA*

Neutron characteristics of salt blanket micromodels containing eutectic mixtures of sodium, zirconium and uranium sulphides were measured on FKBN-2M, BIGR and MAKET installations. The effective fission cross sections of neptunium, plutonium, americium and curium isotopes were measured on the neutron spectra formed by micromodels.

*KEYWORDS: transmutation, minor actinides, fluoride salts, micromodel, critical assembly, neutron spectrum, multiplication coefficient, fission, effective cross section, nuclear track detector, nuclear data library*

## I. Introduction

The additives of minor actinides may constitute a considerable part of the core in the nuclear transmutation facilities under development. The requirements to the accuracy of nuclear data for such isotopes increase essentially. In fast reactors and blankets of facilities aimed at spent nuclear fuel and weapon plutonium transmutation there can be successfully used molten mixtures of fluoride salts. A cycle of neutron characteristics researches for a set of most promising eutectic mixtures of fluoride salts is carried out in VNIIEF and ITEP.

## II. Critical experiments and measurements of neutron spectra carried out in VNIIEF

In VNIIEF there were used two micromodels of salt blanket (MSB). The mole composition of the mixture in MSB-1 is as follows: $0.517NaF+0.483ZrF_4$; while the one of MSB-2 is the following: $0.505NaF+0.467ZrF_4+0.028UF_4$. Enrichment of uranium by $^{235}U$ is 90%.

Each salt blanket model is comprised of two hemispheres. Bodily MSB is a sphere 18.3 cm in diameter, its axial channel 2.3 cm in diameter passing through the pole of one of hemispheres. The channel is designed to place neutron detectors to the center of the model. The bodies of the models are produced of sheet zirconium 0.1cm thick. The weight of fluoride powders and total mass of each model (in terms of zirconium shell) constituted 8354 g and 9306 g – for MSB-1; 8411 g and 9367 g – for MSB-2.

As fast neutron sources there were used in VNIIEF critical assemblies (CA) that were installed on a special facility – FKBN-2M (Russian abbreviation for "Physical Reactor of Fast Neutrons - Second Modification").

The critical assemblies (CA) are assembled of components of hemispheric shape. Bodily a critical assembly has the form close to spherical. MSB is established in the internal cavity of CA (see **Fig. 1**). The spherical layers of metal uranium enriched by $^{235}U$ isotope up to ~90% (three layers) and up to ~36% (three layers as well) are placed onwards on the radius. All the components of metal uranium are covered with anticorrosive copper-nickel layer ~50 μ thick. In the experiments there were used two critical assemblies: the first assembly – in the experiments with MSB-1, the second one – in the experiments with MSB-2. The parameters of critical assemblies are given in **Table 1**. Before arrangement each critical assembly is separated to two knowingly sub-critical parts. The lower part can be displaced vertically, while the upper one – horizontally (see **Fig. 2**). The continuous control over the assembly reactivity was implemented through measuring the density of the flux of leakage neutrons. A detailed description of the technique aimed at measuring the CA characteristics is given in paper[1].

---

* Corresponding author, Tel. 83-130-45989, Fax. 83-130-45569,
  E-mail: fomushkin@expd.vniief.ru

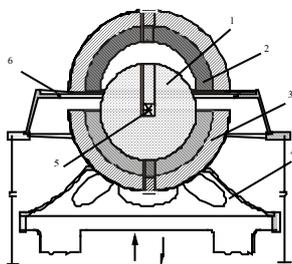

**Fig. 1** Critical assemblies containing salt blanket models used in the experiments on FKBN-2M (RFNC-VNIIEF)
1 – salt blanket model; 2 – layer of $^{235}$U(90%); 3 – layer of $^{235}$U(36%); 4 – lower block support; 5 – neutron detector; 6 – upper block support

In the measurements with MSB-1 and MSB-2 there were defined the following parameters: $K_{eff}$ – for the assemblies with salt blanket micromodels occurring in the central cavity, $\Delta K_{eff}$ – reactivity disturbance conditioned by the MSB removal from the central cavity, constants of prompt neutrons density decrease for two critical systems. The results of critical parameters measurement are presented in Table 1.

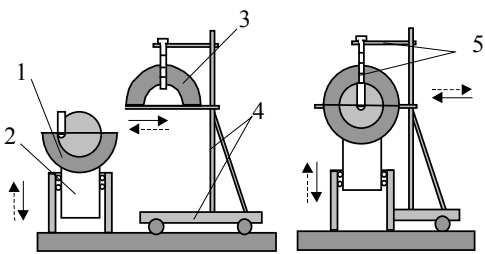

**Fig. 2** Sequence of operations at realizing critical experiments on FKBN-2M facility
1 – CA lower block; 2 - vertically removed support; 3 – CA upper block; 4 – horizontally removed support; 5 – mechanism aimed at detectors loading

To define the characteristics of the neutron field in the MSB center there were measured fission and activation integrals of the following nuclides: $^{235}$U; $^{238}$U; $^{237}$Np; $^{239}$Pu; $^{240}$Pu; $^{197}$Au(n,γ)$^{198}$Au; $^{64}$Cu(n,γ)$^{65}$Cu; $^{115}$In(n,n′)$^{115m}$In; $^{58}$Ni(n,ρ)$^{58}$Co; $^{27}$Al(n,α)$^{24}$Na; $^{32}$S(n,ρ)$^{32}$P. The neutron spectra obtained as a result of integral experiments processing are given on **Fig. 3**.
Besides the experiments on FKBN-2M facility a considerable scope of measurements was performed under external irradiation of MSB-1 and MSB-2 by the neutrons of uranium-graphite reactor BIGR. There were measured neutron spectra (**Fig. 4**) and effective fission cross sections of neptunium, plutonium, americium and curium isotopes (see below).

The neutron spectra of MSB-1 and MSB-2 contain a peak of near-thermal-energy neutrons with $E_n \cong 0.35$ eV maximum, and moderation spectrum $\sim 1/E_n$; the spectrum of fast neutrons was approximated by a sum of two rows by Chebyshev-Lagguerre polynomial with the shape parameters $\alpha$=1/2 and 1. The neutron temperatures of the corresponding components are T$\cong$0.9 MeV and 0.03 MeV. When FKBN-2M or BIGR are used as sources of primary neutrons the contributions of separate components to the spectra of MSB are different. On the other hand, when working with one and the same source of primary neutrons the spectra of MSB-1 and MSB-2 practically coincide (see Fig. 3, 4).

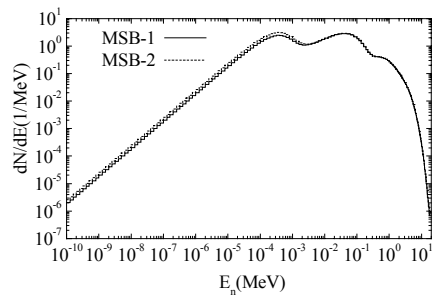

**Fig. 3** Neutron spectra in the center of MSB (FKBN-2M)

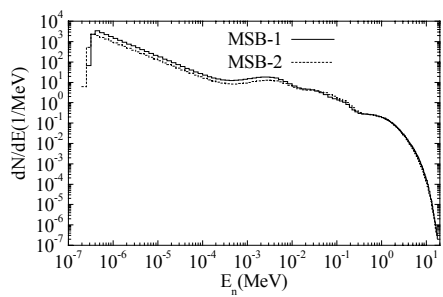

**Fig. 4** Neutron spectra in the center of MSB (BIGR)

At measuring efficient fission cross sections the fission cross section of $^{235}$U averaged by the corresponding neutron spectrum was used as a reference one. At calculating the values $<\sigma_f(^{235}U)>$ the data from TENDL[2] were used.

### III. Critical Experiments and Neutron Spectra Measurements Performed in ITEP

A micromodel used in ITEP is of cylindrical shape 230 mm in diameter and 500 mm high, its body being produced of plutonium. The model is filled with salt eutectic fusion cake 0.52NaF+0.48ZrF$_4$. The irradiated samples are installed in cylindrical channels at different distances (0 mm, 46.5 mm and 96.5 mm) from the model axis. The central channel 58 mm in diameter is filled with salt mixture or a fuel channel (FC) of heavy water reactor MAKET is introduced to it. Fuel channels form in the reactor a hexagonal lattice of 100-mm spacing. The salt micromodel is installed in the center of the assembly on a 400-mm height from the lower remoting lattice.

To carry out the experiments there were formed two hexagonal fuel lattices. The first lattice contained 34 FC and 67272 g of salt mixture (salt insert in the central channel). The second one contained 33 FC and 655592 g of salt mixture (8 fuel elements in the central channel). All FCs were assembled of fuel elements of bushing type, each of them contained uranium of 90% enrichment by $^{235}$U.

To evaluate the neutron spectra in the channels of MSB-ITEP micromodel there were measured the following reaction rates: $^{235}$U(n,f), $^{238}$U(n,γ), $^{55}$Mn(n,γ), $^{63}$Cu(n,γ), $^{176}$Lu(n,γ), $^{197}$Au(n,γ), $^{115}$In(n,n′), $^{27}$Al(n,α), $^{64}$Zn(n,p). The neutron spectra in the channels obtained on the basis of the measured reaction rates are presented on **Fig. 5**. The neutron spectra in the channels are close to the thermal spectrum. The share of the neutron spectrum with $E_n < 0.1$ eV varies from 40% (when the central channel of MSB is filled with salt mixture) to 31% (when the fuel element of MAKET reactor is installed in the central channel). Moreover, the share of fast neutrons ($E_n > 0.1$ MeV) varies from 5 to 15%, correspondingly.

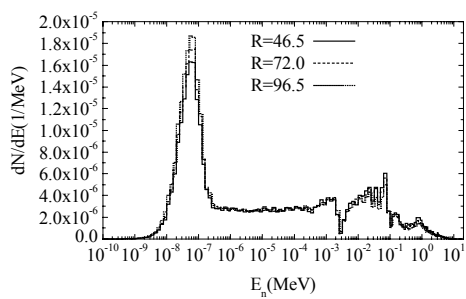

**Fig. 5** Neutron spectra at different distances from MSB "MAKET" (lattice 2)

Depending on the lattice type and on the distance from MSB ITEP axis to the corresponding channels the efficient fission cross section of $^{235}$U lies within the following limits: from 160.4±4.0 barn (lattice 2, R=46.5 mm) to 229.9±5.7 barn (lattice 2, R=96.5 mm).

In accord with the estimates the disturbance introduced by MSB ITEP micromodel to the reactivity of MAKET reactor constitutes $\Delta K_{eff}$=-0.0580 for the first lattice and -0.0669 for the second one.

## IV. Measurements of Efficient Fission Cross Sections by the Neutrons of Salt Blanket Micromodels

The measurements of efficient fission cross sections on VNIIEF and ITEP installations were performed using one and the same technique developed and widely used in VNIIEF.

Highly enriched samples of uranium, plutonium, americium and curium isotopes were obtained using SM-2 electromagnetic mass-separator. Sector magnetic field $H=H_0 \cdot R_0/R$ with the average trajectory radius $R_0$=1000 mm and 2-radian deflection angle is used in the separator.

The mass-spectrometric analysis of enriched samples was realized with the accuracy not worse than 0.1% (at relative content of admixtural isotope $\geq 10^{-2}$) and with the accuracy of 1-3% (at $<10^{-4}$ content of admixture). For the majority of samples the enrichment by the basic isotope was not worse than 98%; the data are given in **Table 2**.

After chemical purification the fissionable material was applied on substrates using a method of nitrates electrolysis from aqueous or alcoholic solutions. The spot diameter in all produced layer-targets is 6 mm. The uniformity of active material distribution over the spot was controlled using a method of autoradiography.

To weigh the samples, i.e. to determine the quantity of active nuclei in the layers there were used α- and γ-spectrometry methods as well as spontaneous fission and fission of nuclei by thermal neutrons. In α- and γ-spectrometers there were applied semiconductor detectors with a standard set of electronic equipment. For "weighing" by spontaneous fission and fission by thermal neutrons there was used a technique identical to the technique of basic measurements, i.e. with the use of dielectric track detectors.

Dielectric track detectors – silicate glass and film of polycarbonate with 90000 molecular mass – were used in the measurements. These detectors are insensible to α-, β-, γ-, n-radiation and under the corresponding conditions they register fission fragments with 100% efficiency.

Glass or polymer plates were located in special containers – measurement chambers – in parallel and coaxially with the layer of fissionable isotope at 6-mm distance from the layer. A circle 6 mm in diameter (registration area) was marked out on the detector surface by a special diaphragm. The accuracy of all dimensions in measuring chambers was controlled with the aid of instrumental microscope.

The chamber geometric efficiency Ω or the probability of a fragment hit to the registration area was calculated basing on the supposition that fissionable material is uniformly distributed over the layer. The effective number of active nuclei in each layer $\langle n \cdot \Omega \rangle$, i.e. the full number of nuclei multiplied by the chamber efficiency was used at processing of the results. It should be mentioned that at "weighing" the layers by spontaneous fission or by nuclei fission by thermal neutrons the $\langle n \cdot \Omega \rangle$ value was measured immediately, moreover, the possible non-uniformity of the material distribution over the layer was taken into account automatically.

At each measurement two chambers were placed to the central channel of the salt blanket micromodel: the first chamber – with the layer of investigated isotope, the second one – with a benchmark layer of $^{235}$U or $^{239}$Pu. The parameters of benchmark layers are also presented in the Table 2. Each isotope under investigation was exposed to irradiation at least four times: two times with uranium benchmark and two times – with plutonium one.

The duration of track detector occurrence in the measuring chamber was under strict control so that the background of nuclei spontaneous fission is taken into account afterwards. As a rule this time was as long as 10-11 minutes. In separate series there was measured such background of nu clei spontaneous fission that is of particular significance for even isotopes of curium.

Irradiation events were performed under atmospheric pressure, i.e. with no evacuation of chambers. Preliminary researches demonstrated that at the selected geometric dimensions the influence of atmospheric air to the efficiency of fragment registration and especially to the final result of measurements is negligibly low.

After irradiation the glass detectors were extracted from chambers and underwent chemical etching. The review of detectors and the count of the number of tracks in each detector were implemented with the aid of optical microscope. Each detector was reviewed independently by two highly qualified operators. The divergences between their readings did not as a rule exceed ~0.5%.

In the course of processing there were taken into account the contribution of admixtural isotopes to the total number of registered fission fragments and the background of spontaneous fission. The effective (average) fission cross sections of $^{235}$U for the neutron spectra used are presented in Table 2; the data on $\sigma_f(E)$ for $^{235}$U were taken from TENDL library[2]. The presented values $<\sigma_f(^{235}U)>$ were used in the fission cross section measurements of transuranium isotopes as benchmark ones. The results of measuring $<\sigma_{fi}>/<\sigma_f(^{235}U)>$ relations are given in Table 2.

Because of the lack of space in Table 2 not all the results obtained on MAKET reactor are presented. The preliminary analysis of experimental data demonstrated that considerable divergences with calculation values $<\sigma_{fi}>$ were not registered. The most unsatisfactory agreement between the calculation and experimental data was observed for $^{240}$Pu, $^{247}$Cm (MAKET reactor measurements).

The work has been performed under the ISTC Project No1145 entitled "Nuclear-Physics Researches Aimed at Solving the Problems of Weapon Plutonium Conversion and Long-Lived Radioactive Wastes Transmutation" supported by JAERI (Japan). In part, the work has been supported by the U.S. Department of Energy.

**Table 1** Characteristics of critical ($K_{eff}$=1±0,0025) spherical assemblies containing salt blanket models

| Blanket model | Spherical layer of $^{235}$U(90%) $R_{int.}$=9.15cm $R_{ext.}$=12.25cm Mass, g | Spherical layer of $^{235}$U(90%) $R_{int.}$=12.25cm $R_{ext.}$=15.00cm Mass, g | $\Delta K_{eff}$, introduced by MSB | $\alpha$, μs$^{-1}$ |
|---|---|---|---|---|
| MSB-1 | 80154 | 116314 | +0.0169 | 0.39 |
| MSB-2 | 79663 | 115881 | +0.0226 | 0.45 |

$\alpha$ - constant of prompt neutrons density decrease; for CA without MSB $\alpha$=0.73 μs$^{-1}$.

**Table 2** Results of effective fission cross section measurements

| Isotope | Enrichment (%) | BIGR MSB-1 $\sigma_f(^{235}U)$=2.5389 b $\sigma_{fi}/\sigma_f(^{235}U)$ | BIGR MSB-2 $\sigma_f(^{235}U)$=2.2058 b $\sigma_{fi}/\sigma_f(^{235}U)$ | FKBN-2M MSB-1 $\sigma_f(^{235}U)$=1.4488 b $\sigma_{fi}/\sigma_f(^{235}U)$ | FKBN-2M MSB-2 $\sigma_f(^{235}U)$=1.4597 b $\sigma_{fi}/\sigma_f(^{235}U)$ | MAKET Latice 1, R=0 $\sigma_f(^{235}U)$=218.3 b $\sigma_{fi}/\sigma_f(^{235}U)$ | MAKET Latice 2, R=46.5 $\sigma_f(^{235}U)$=160.4 b $\sigma_{fi}/\sigma_f(^{235}U)$ |
|---|---|---|---|---|---|---|---|
| $^{235}$U | 99.974 | 1.000(31) | 1.000(31) | 1.000(31) | 1.000(31) | 1.000(36) | 1.000(37) |
| $^{238}$U | 99.99 | | | 0.072(03) | 0.074(03) | | |
| $^{239}$Pu | 99.99 | 1.149(37) | 1.122(36) | 1.205(38) | 1.214(39) | 1.431(51) | 1.693(61) |
| $^{237}$Np | 99.99 | | | 0.514(17) | 0.524(17) | 0.037(02) | 0.0140(01) |
| $^{238}$Pu | 99.16 | 0.557(19) | 0.482(20) | | | 0.0362(36) | 0.0429(3) |
| $^{239}$Pu | 99.99 | 1.149(37) | 1.122(36) | 1.205(38) | 1.214(39) | 1.431(51) | 1.693(61) |
| $^{240}$Pu | 99.517 | 0.242(11) | 0.246(12) | 0.486(21) | 0.503(23) | 0.550(22) | 0.153(07) |
| $^{241}$Pu | 99.995 | 1.381(80) | 1.151(82) | | | 2.197(94) | 1.578(68) |
| $^{242}$Pu | 99.985 | 0.193(70) | 0.218(08) | | | | |
| $^{244}$Pu | 84.48 | 0.236(17) | 0.242(17) | | | | |
| $^{241}$Am | 99.999 | 0.267(09) | 0.260(09) | 0.544(18) | 0.542(17) | | |
| $^{242m}$Am | 84.37 | 3.144(179) | 2.765(157) | 2.132(111) | 2.135(107) | 13.58(80) | 12.91(68) |
| $^{243}$Am | 99.99 | 0.175(08) | 0.205(09) | | | | |
| $^{243}$Cm | 98.237 | 2.433(105) | 2.242(101) | 2.009(86) | 1.868(84) | 1.232(60) | 1.514(72) |
| $^{244}$Cm | 99.58 | 0.282(12) | | | | | |
| $^{245}$Cm | 98.96 | 1.570(69) | 1.479(67) | 1.521(65) | 1.541(67) | 9.268(491) | 3.463(126) |
| $^{246}$Cm | 99.87 | 0.257(12) | | | | | |
| $^{247}$Cm | 83.20 | 1.568(80) | 1.027(55) | | | 0.467(28) | 0.529(32) |
| $^{248}$Cm | 97.44 | 0.246(13) | | | | | |